\documentclass[12pt]{article}

\usepackage{amsmath}
\usepackage{amssymb}
\usepackage{amsfonts}
\usepackage{mathrsfs}
\usepackage{graphicx}
\usepackage{enumerate}
\usepackage{subcaption}

\usepackage[margin=1in]{geometry}
\usepackage[width=0.9\textwidth]{caption}

\usepackage[pdftex,colorlinks=true,linkcolor=blue,citecolor=blue,urlcolor=blue]{hyperref}

\begin{document}

\title{Complexity and typical microstates}
\author{Simon F. Ross\footnote{s.f.ross@durham.ac.uk} \\  \bigskip \\ Centre for Particle Theory, Department of Mathematical Sciences \\ Durham University\\ South Road, Durham DH1 3LE}

\maketitle
 
\begin{abstract}
Typical black hole microstates in AdS/CFT were recently conjectured to have a geometrical dual with a smooth horizon and a portion of a second asymptotic region. I consider the application of the holographic complexity conjectures to this geometry. The holographic calculation leads to divergent values for the complexity; I argue that this classical divergence is consistent with expectations for typical microstates. 
\end{abstract}
 
\clearpage 

\section{Introduction}
 
The AdS/CFT correspondence provides a non-perturbative definition of quantum gravity on asymptotically anti-de Sitter spaces, including black holes in the bulk spacetime. An essential element in the holographic dictionary is understanding the description of black holes in this correspondence, particularly the region behind the horizon. This is understood for the eternal black hole, which is dual to the thermofield double state, a particular entangled state in two copies of the CFT \cite{Maldacena:2001kr}. This state reduces to a thermal density matrix in each copy of the CFT, which is related to the region outside the horizon. Reconstruction of the region inside the horizon requires degrees of freedom from both copies of the CFT. In particular, approximate bulk field operators in the region behind the horizon can be reconstructed from the dual local operators in the two copies of the CFT \cite{Papadodimas:2012aq}.

In a typical high energy pure state in a single copy of the CFT, simple observables will be close to thermal values; this is due to entanglement between these coarse-grained features of the state and some fine-grained details that the simple observables do not measure. It is thus widely believed that a typical high energy pure state has a bulk description as an AdS-Schwarzschild black hole, which includes at least the region outside the horizon. 

The question of whether there is a region behind the horizon is controversial. It has been argued that the resolution of the information loss paradox requires a breakdown of the geometric description at the horizon scale for typical states \cite{Mathur:2009hf,Almheiri:2012rt,Almheiri:2013hfa}. Papadodimas and Raju have argued by contrast that the entanglement of the coarse-grained and fine-grained features of the typical state has the same general character as the entanglement between the two copies of the CFT in the thermofield double state, so one could expect to be able to recover a region behind the horizon in this case just as in the thermofield double state \cite{Papadodimas:2012aq,Papadodimas:2013jku,Papadodimas:2015jra}. They have used Tomita-Takesaki theory to construct state-dependent ``mirror operators'' which are entangled with the local operators in the CFT.   They constructed candidate bulk field operators in the region behind the horizon using both the local CFT operators and the mirror operators, giving a bulk geometry with a smooth horizon for typical states. The state-dependent nature of this construction is supposed to avoid the arguments in favour of the breakdown of the geometric description at the horizon.

\begin{figure}
\centering
		 \includegraphics[width = 0.4 \textwidth]{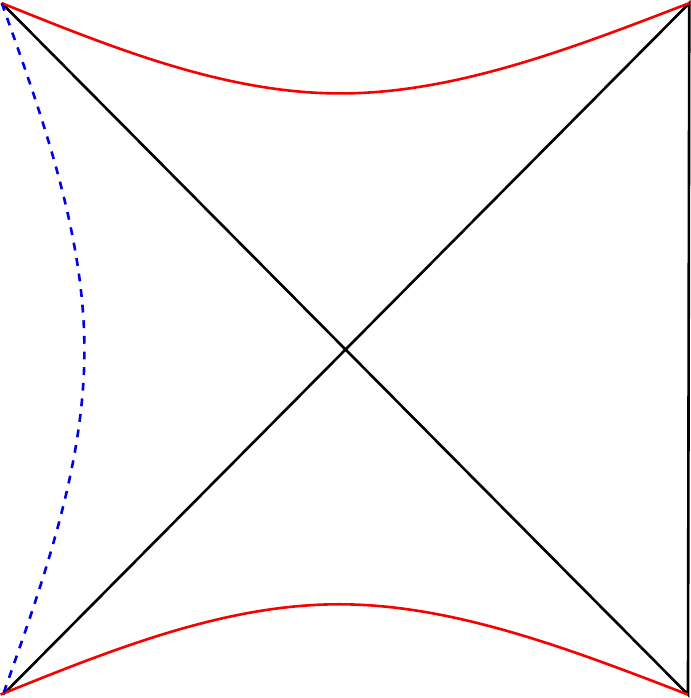}
\caption{The proposed geometry dual to a typical high energy state in a CFT. The CFT lives on the boundary on the right; the dotted line on the left indicates a limit of the geometrical description of the state.}
\label{geom}
\end{figure}

Recently, \cite{deBoer:2018ibj,deBoer:2019kyr} argued that the geometry recovered in this way also includes a part of the other asymptotic region of the eternal black hole spacetime, as depicted in figure \ref{geom}. This picture was motivated by the effective time-independence of the typical high-energy state, which was argued to imply an approximate Killing symmetry of the dual geometry. The local bulk fields in the left region are reconstructed purely from the mirror operators. Since the dual of this geometry is a typical pure state in a single CFT, the Penrose diagram cannot be extended arbitrarily to the left. The dotted line indicates a breakdown of the geometric description. Related constructions for atypical states were given in \cite{Kourkoulou:2017zaj,Almheiri:2018ijj,Almheiri:2018xdw,Cooper:2018cmb}. 

The aim of the present paper is to test this proposal by considering the application of the holographic complexity conjectures to the geometry in figure \ref{geom}, and comparing to the expectations for the complexity of a typical state. The holographic complexity conjectures \cite{Susskind:2014rva,Stanford:2014jda,Susskind:2014jwa,Susskind:2014moa,Brown:2015bva,Brown:2015lvg} relate the complexity of the CFT state to features of the bulk geometry. In the complexity-volume (CV) conjecture, the complexity of the state on a spacelike slice $\Sigma$ of the boundary is related to the volume of the maximal volume slice $B$ in the bulk whose boundary is $\Sigma$. In the complexity-action (CA) conjecture, the complexity is instead related to the bulk action evaluated on the ``Wheeler-de Witt'' (WdW) patch in the bulk.

I will describe the application of these conjectures to the bulk geometry in figure \ref{geom}. They produce divergent results, which I argue is consistent with the expectations for the complexity of the typical state: the complexity should be exponentially large (of order $e^S$ where $S$ is the entropy of the ensemble the state is chosen from), and the description of the typical state by the geometry in figure \ref{geom} treats such an exponential as infinity. In particular, the time independence of the typical state used to argue for this geometry is only true up to corrections of order $e^{-S}$, so there is no reason to expect this geometry to be valid when we consider time scales of order $e^S$. Corrections to the geometry at these time scales could cut off the divergence to produce the expected order $e^S$ value of the complexity of the typical state. Thus this calculation provides some evidence in favour of the proposed geometry of figure \ref{geom}. 

In section \ref{vol}, I consider the application of the CV conjecture; see figure \ref{comp}. There is a close connection between the divergence of the complexity in this geometry and the late-time growth of the complexity in a black hole formed by gravitational collapse, or in the time evolution of the thermofield double state. This also supports the proposal that this geometry provides a description of the typical state. For the CV calculation, the coefficient of the divergence precisely agrees with the late time growth rate of the complexity. The calculation depends only on the region behind the horizon and does not probe the left asymptotic region, so this calculation is in this sense not a very strong test of the proposal of \cite{deBoer:2018ibj,deBoer:2019kyr}.

In section \ref{act}, I consider the application of the CA conjecture. We will see that there is a subtlety in the application of the CA conjecture to this geometry, in deciding what we want to take the WdW patch to be, see figure \ref{wdw}. The simplest choice is to take the WdW patch to be the domain of dependence of a bulk Cauchy surface ending on the boundary surface $\Sigma$; this gives us the WdW patch in the right picture in figure \ref{wdw}. However, in this geometry, the maximal-volume surface $B$ considered in the CV conjecture is not a Cauchy surface for the full bulk spacetime, as it does not extend into the left region, as depicted in figure \ref{comp}. We could therefore take the WdW patch to be the domain of dependence of the surface $B$; this gives us the smaller region in the left picture of figure \ref{wdw}. 

If we take the WdW patch to be the domain of dependence of $B$, there is again a match between the divergence and the growth rate in the time-dependent scenario. If we take the domain of dependence of a Cauchy slice, the coefficient of the divergence is different. This gives some reason for preferring the former definition, although perhaps we should not necessarily expect too straightforward a relation between the typical state geometry and the time-dependent case. On the other hand, the former calculation depends only on the region behind the horizon and does not probe the left asymptotic region. The second calculation, by contrast, depends on the region to the left, and the coefficient of the divergence will depend on where we put the cutoff on the left and what assumptions we make about the physics of this cutoff. I will argue that for this calculation, it appears natural to have this cutoff at a radial position $r_L$ of order the horizon radius. There is some tension with the arguments in  \cite{deBoer:2018ibj,deBoer:2019kyr}, who favoured a large value for the cutoff. 

\section{CV conjecture}
\label{vol} 
 
In the CV conjecture \cite{Susskind:2014rva}, the complexity $\mathcal C$ of a pure state $|\Psi \rangle$ of a holographic field theory on some spatial slice $\Sigma$ on the boundary of an asymptotically AdS spacetime is identified with the volume $V$ of the maximal volume codimension one slice $B$ in the bulk having its boundary on $\Sigma$,
\begin{equation}
{\mathcal C_{\textnormal{V}}} \propto \frac{V(B)}{G_{\textnormal{N}} l_{\textnormal{AdS}}}.
\end{equation}
This was motivated by the study of the behaviour of Schwarzschild-AdS black hole solutions, where it was found that the volume of the maximal volume slice grows linearly with time, even at late boundary times when other observables have thermalized. The volume grows linearly because the slice approaches a constant-$r$ surface of maximal volume in the region inside the black hole, which we call $B_{in}$; as time increases the surface $B$ is close to $B_{in}$ over a larger range of $t$. 
 
\begin{figure}
\centering
	 \includegraphics[width = 0.4 \textwidth]{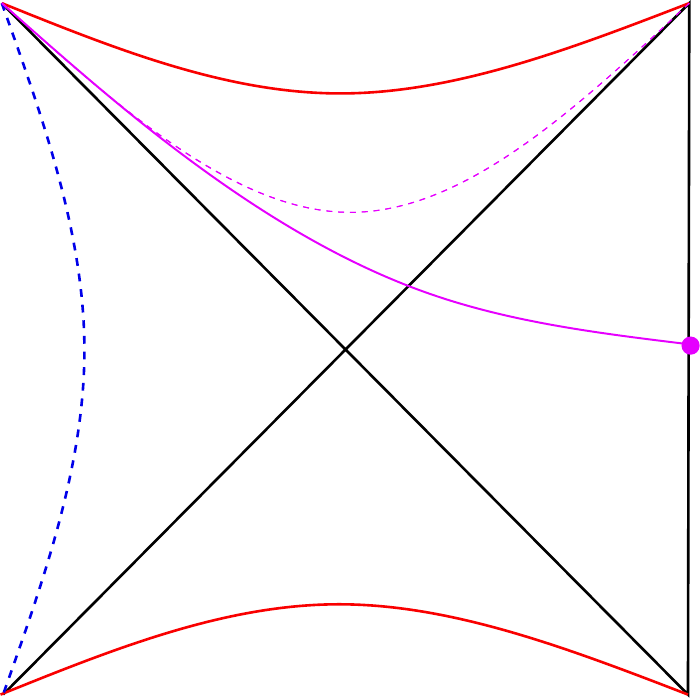}
\caption{In the CV calculation, the complexity is given by the divergent volume of the slice $B$, which approaches $B_{in}$ in the left part of the diagram.}
\label{comp}
\end{figure}

When we apply this to the geometry of figure \ref{geom}, we get a divergence, as the surface $B$ is constrained only to approach the boundary slice $\Sigma$ (indicated by a large dot  in figure \ref{comp}) and is otherwise free to vary in the bulk. We can obtain infinite volume by allowing $B$ to asymptotically approach $B_{in}$ in the left side, as illustrated in figure \ref{comp}. Note that the geometry is time-symmetric, so we can take as the maximal-volume slice either the surface drawn in figure \ref{comp} in the black hole region to the future or its mirror image in the white hole region to the past. 

This argument is independent of the details of the geometry, but to be more explicit, take the black hole to be the AdS-Schwarzschild solution in $d+1$ bulk dimensions, with $d \geq 3$. The bulk metric is 
 \begin{equation} \label{schw}
 ds^2 = -f(r) dt^2 + \frac{dr^2}{f(r)} + r^2 d\Sigma_{d-1}^2,  
 \end{equation}
where $f(r) = \frac{r^2}{\ell^2} + k - \frac{\mu}{r^{d-2}}$,  and $\ell$ is the AdS length scale, $k=0,\pm1$ and $d\Sigma_{d-1}^2$ is the metric on a unit sphere $S^{d-1}$ for $k=1$, a flat plane for $k=0$, and the hyperbolic space $H^{d-1}$ for $k=-1$. The mass of the black hole is $M = \frac{(d-1) \Omega_{d-1}}{16 \pi G} \mu$, where $\Omega_{d-1}$ is the  volume of the space of constant $t$ and $r$ with respect to the metric $d \Sigma^2_{d-1}$  (for $k=0,-1$, we consider a compactification to make this volume finite). The solution has a horizon at $r_+$ where $f(r_+)=0$. 

The maximal constant-$r$ surface $B_{in}$ inside the horizon is at the radius $r= r_{in} < r_+$ where $\partial_r ( f(r) r^{2d-2}) = 0$. The volume of $B_{in}$ is 
 \begin{equation}  \label{vin} 
 V_{B_{in}} = \sqrt{ - f(r_{in}) } r_{in}^{d-1} \Omega_{d-1} \int dt, 
 \end{equation}
This is divergent because of the integral over $t$. 

Consider the state of the CFT on a surface $\Sigma$ at $t=0$ on a cutoff boundary $r=r_{max}$ (the cutoff is introduced to regulate $UV$ divergences).  We consider the maximal volume slice which moves away from $\Sigma$ towards $t>0$, crosses the future horizon and approaches the surface $B_{in}$ in the future black hole region. It will approach $B_{in}$ as $t \to -\infty$,\footnote{Defining a time coordinate in the future region which increases to the right, by continuity of the Killing vector across the right horizon.} and the volume will hence diverge from integrating over large negative times. 

The divergence in this volume is essentially the same as the unbounded growth of the volume of the maximal volume slice in the time evolution in a collapse geometry or the thermofield double state. In these dynamical situations, the maximal volume slice will lie along $B_{in}$ for a finite range of times, $\Delta t \approx t$ for late times, producing a linear growth of complexity with $t$. The growth rate of the complexity in the dynamical calculations is thus simply the volume of $B_{in}$ per unit time in \eqref{vin}, which similarly gives the coefficient of the divergence in the present calculation. Thus, the holographic complexity we obtain  for typical states from the proposed geometry is essentially the infinite time limit of the holographic complexity in the dynamical cases.

In both the dynamical situation and our consideration of the typical state, the divergence is an artifact of the classical description. We expect this classical geometrical picture to break down when we consider observations over time scales of order $e^S$. In the typical state case, this can be seen quite simply from the  arguments in  \cite{deBoer:2018ibj,deBoer:2019kyr}: this geometry was proposed because typical states look approximately time-independent when probed by simple observables: 
\begin{equation} \label{corr}
\langle \psi | \frac{d \mathcal O}{dt} | \psi \rangle = \mathrm{Tr}[ \rho_m \frac{d \mathcal O}{dt} ] + O(e^{-S}) = O(e^{-S}), 
\end{equation}
where the first term in the second equality vanishes by the time-independence of the microcanonical density matrix $\rho_m$, and $S$ is the entropy of the ensemble. The $O(e^{-S})$ corrections could lead to breakdowns of the time-independent geometrical description on timescales of order $e^S$. The complexity of states in the quantum mechanical system is bounded above roughly by $e^S$, as this is the dimension of the space of states we are considering. We do not have control over the nature of the corrections in \eqref{corr}, but the scaling is right for them to cut off the classical divergence found in this calculation to reproduce the expected maximum complexity. 

\section{CA conjecture}
\label{act}

In the  CA conjecture \cite{Brown:2015bva,Brown:2015lvg},  the complexity of $|\Psi \rangle$ is identified with the action of the  WdW patch. The proposal is that
\begin{equation} \label{ca}
{\mathcal C_{\textnormal{A}}} = \frac{I}{\pi \hbar},
\end{equation}
where $I$ is the action of the WdW patch. This proposal has the advantage that the formula is more universal, containing no explicit reference to a bulk length scale. It is also often easier to calculate, as there is no maximisation problem to solve. Finding the WdW patch for a given boundary slice is easier than finding the maximal volume slice. We adopt the action prescription of  \cite{Lehner:2016vdi}, with the additional counterterm required for parametrization-independence on the null boundaries. The action of the WdW patch also exhibits linear growth in time at late times for the black holes, saturating a conjectured  universal upper bound on the rate of growth of the complexity \cite{lloyd}
 \begin{equation} \label{ctd}
 \frac{d \mathcal C}{dt} \leq \frac{2 M}{\pi \hbar}. 
 \end{equation}

\begin{figure}
\centering
	\begin{subfigure}[b]{.475\textwidth}\centering
	 \includegraphics[width = 0.6 \textwidth]{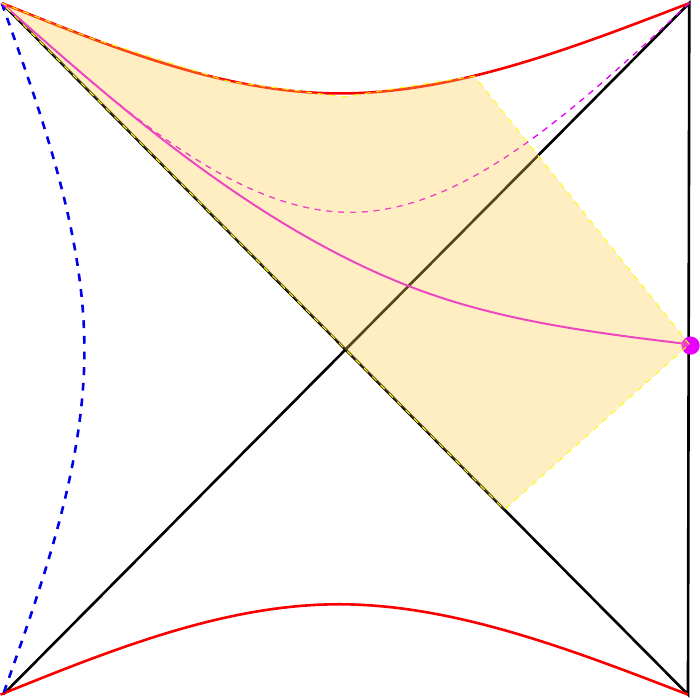}
	\end{subfigure}
	\begin{subfigure}[b]{.475\textwidth}\centering
	 \includegraphics[width = 0.6 \textwidth]{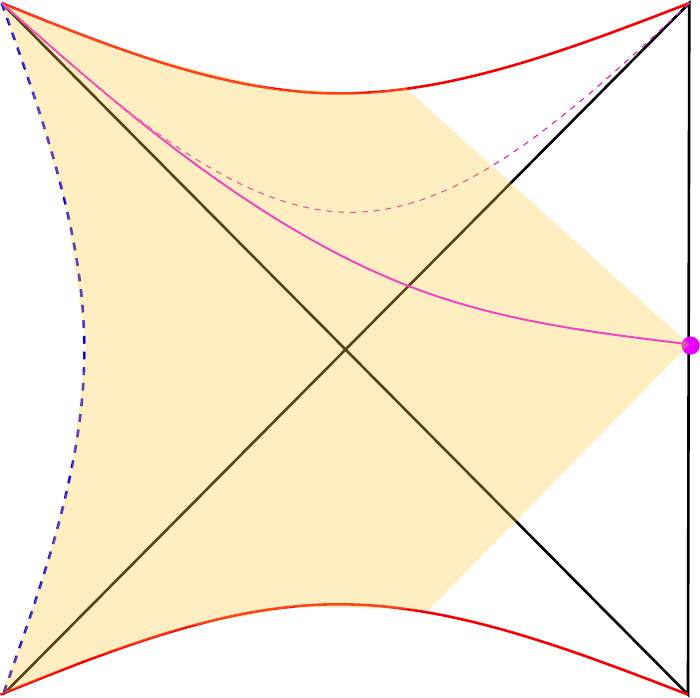}
	\end{subfigure}
\caption{Two alternatives for the Wheeler-de Witt patch: on the left, the domain of development of $B$, and on the right, the domain of development of a Cauchy surface.}
\label{wdw}
\end{figure}

For the CA conjecture, there is a subtlety, as there are two possible choices for the WdW patch. We usually take the WdW patch to be the domain of development of the surface $B$. In the examples considered to date, however, this surface was also a Cauchy surface for the bulk spacetime. In our present example, the surface $B$ we considered above is not a Cauchy surface for the full spacetime, as it never enters the left asymptotic region.\footnote{Note that this is a subtle issue, and depends sensitively on exactly how we define $B$. We could start by requiring the bulk slice to be a Cauchy surface, extending all the way to the dotted line on the left in figure \ref{geom}, and try to extremise its volume in this family of Cauchy surfaces;  this constrained minimization problem produces the same divergence as before, as the surface lies increasingly along $B_{in}$ as we move to the future.} So we have a choice: either we consider the domain of development of $B$, as depicted on the left in figure \ref{wdw}, or we consider the domain of development of a Cauchy surface, as depicted on the right in figure \ref{wdw}.\footnote{In drawing the latter WdW patch, we have assumed that there are no independent boundary conditions we need to impose at the left boundary, so knowledge of the state on a Cauchy surface is sufficient to determine evolution along the boundary. This seems appropriate as this geometry is supposed to correspond to a state in the CFT defined on the right boundary.}  Both these choices have their attractions: the former will produce results which accord nicely with the late time limit of the time dependent cases, while the latter seems more natural and offers a more useful probe of the full geometry.

We will first discuss the action calculation in the former case, where we take the domain of dependence of $B$. To do the calculation, we cut off $B$ at some large time $-t_{max}$ on the left. The calculation of the action for this patch is very similar to the calculation for the eternal two-sided black hole in \cite{Lehner:2016vdi,Carmi:2017jqz} (mapped to a coordinate system with $t_L=t, t_R=0$). The difference is that the WdW patch in their case is extended into the left asymptotic region. But at large $t$, the past null boundary in the left region approaches the horizon, while the past null sheet from $t = -t_{max} ,r = r_{in}$ in our case approaches the horizon from the inside at large $t_{max}$. As a result, as we will explicitly see below, the divergence in this calculation will have a coefficient which is the same as the late time rate of growth with $t$ in the eternal black hole. 

The action of the Wheeler-de Witt patch in the prescription of \cite{Lehner:2016vdi} is 
\begin{eqnarray} \label{adsa}
I &=& \frac{1}{16\pi G} \int_{W}d^{d+1}x \sqrt{-g} \,  (R- 2 \Lambda)     + \frac{1}{8\pi G} \int_S d^d x  \sqrt{-h}   \, K \\ &&- \frac{1}{8\pi G}   \int_{N} d^{d-1} x d\lambda \, \kappa  - \frac{1}{8\pi G}  \int_{N}    d^{d-1} x d\lambda \, \Theta \ln | \ell \Theta |   -  \frac{1}{8\pi G}   \int_{\Sigma} d^{d-1}x \sqrt{\gamma} \, a, \nonumber
\end{eqnarray}
where $W$ denotes the WdW patch, $S$ the spacelike boundary at $r = \epsilon$, $N$ the null boundaries, and $\Sigma$ the joints between boundaries. The spacelike boundary contribution is the usual Gibbons-Hawking-York (GHY) term, where $K$ is the trace of the extrinsic curvature. In the null boundary contributions, $\lambda$ is a parameter on the null generators, so $k^\alpha = \partial x^\alpha /\partial \lambda$ is the tangent to the generators, with $\kappa$ defined by $k^\alpha \nabla_\alpha k^\beta = \kappa k^\beta$, and $\Theta = \frac{1}{2} \gamma^{-1} \partial_\lambda \gamma$ is the expansion of the null surfaces, where $\gamma$ is the determinant of the metric on the cross-sections of constant $\lambda$. We will always work with affinely parametrised null generators,  so the first term on the null boundaries vanishes. In the joint contribution, $a = \ln | k \cdot k' /2|$ is determined by the inner product of the two tangent vectors. 

The boundaries of our first WdW patch are the future and past null sheets emanating from $t=0, r=r_{max}$ on the boundary, the future and past null sheets emanating from $t = -t_{max} ,r = r_{in}$ inside the black hole, and a segment along the black hole singularity in between the two null sheets, which we cut off at $r = \epsilon$. There are joints at $t=0, r=r_{max}$, at $t = -t_{max} ,r = r_{in}$, at the intersection of the two past null sheets, and at the intersections of the future null sheets with $r = \epsilon$. The null sheets from  $t=0, r=r_{max}$ lie at $t(r) = \pm ( r^*_{max} - r^*) $, where $r^*(r) = \int \frac{dr}{f(r)}$ is the tortoise coordinate, and $r^*_{max}$ is its value at the cutoff surface. The null sheets from  $t = -t_{max} ,r = r_{in}$ similarly lie at $t + t_{max} = \pm (r^*_{in} - r^*)$, but in this case the minus sign is the future sheet and the plus sign is the past sheet. 

We are interested in the contributions to this action which diverge as we take $t_{max} \to \infty$. There will be such a contribution from the volume term and from the spacelike surface term at $r = \epsilon$. The affine parameter on the null boundaries is proportional to $r$, so the null boundary terms are independent of $t_{max}$, and do not contribute to the divergence. The  joint contributions at $r = \epsilon$ and $r= r_{max}$ are similarly independent of $t_{max}$, which leaves just the joint contribution at the intersection of the past null sheets. This approaches the Killing horizon as $t_{max} \to \infty$, producing a divergence as in the calculation of \cite{Carmi:2017jqz}. 

The volume contribution from the region behind the horizon is 
\begin{eqnarray}
I_{bulk}^F &=&  - \frac{d}{8\pi G \ell^2} \Omega_{d-1} \left[   \int_\epsilon^{r_{in}} dr \, r^{d-1}  (t_{max} + r_{max}^* + r_{in}^* - 2 r^* ) + \int_{r_{in}}^{r_+} dr \, r^{d-1}  (t_{max} + r_{max}^* - r_{in}^*)  \right] \nonumber \\ &=& - \frac{r_+^d}{8\pi G \ell^2} t_{max} + \ldots, \
\end{eqnarray}
where we have used $R-2\Lambda = - \frac{2d}{\ell^2}$, and in the second step we have dropped terms that do not diverge with $t_{max}$ and set $\epsilon =0$. For the surface term, the trace of the extrinsic curvature is $K = \frac{1}{2} \partial_r f + \frac{d-1}{r} f$, so
\begin{equation} 
I_{surf}^F = - \frac{\epsilon^{d-1}}{16 \pi G}  \Omega_{d-1} \left(  \partial_r f (\epsilon)+ \frac{d-1}{\epsilon} f(\epsilon) \right)  (t_{max} + r_{max}^* + r_{in}^* - 2 r^*(\epsilon)).  
\end{equation}
For small $\epsilon$, $f(\epsilon) \approx - \frac{\mu}{\epsilon^{d-2}}$, so 
\begin{equation} 
I_{surf}^F= \frac{d \mu}{16 \pi G }  \Omega_{d-1} t_{max} + \ldots  
\end{equation}
The affine parameter on the null sheets is proportional to $r$. Let us take it to be equal to $r$ (with the inclusion of the expansion counterterm, the value of the action is unchanged by this choice) so the tangent vectors are $k = \pm f^{-1} \partial_t + \partial_r$, and at the intersection of the past sheets, $k \cdot k' = 2f$.  The joint term is then 
\begin{equation} 
I_{joint} = -  \frac{d \mu}{8 \pi G }  \Omega_{d-1} r_{jnt}^{d-1} \log| f(r_{jnt})|,
\end{equation}
where $r_{jnt}$ is determined by solving 
\begin{equation} 
- t_{max} + r^*_{in} - r^*_{jnt} = - r_{max} + r^*_{jnt}. 
\end{equation}
As $t_{max} \to \infty$, $r_{jnt}$ approaches the horizon, which implies $f(r_{jnt}) \approx f'(r_+) (r_{jnt} - r_+)$, and $r^*_{jnt} \sim \frac{1}{f'(r_+)} \log (r_{jnt} - r_+)$, so 
\begin{eqnarray} 
I_{joint} &\approx & -  \frac{1}{8 \pi G }  \Omega_{d-1} r_+^{d-1} \log (r_{jnt} - r_+) \approx -  \frac{1}{8 \pi G }  \Omega_{d-1} r_+^{d-1} f'(r_+) r^*_{jnt} \nonumber \\ & = & \frac{1}{16 \pi G }  \Omega_{d-1} r_+^{d-1} f'(r_+) t_{max} + \ldots 
\end{eqnarray}
Putting it all together, we have that the divergence in the action in the limit as $t_{max} \to \infty$ is 
\begin{equation} 
I = 2M t_{max} + \ldots
\end{equation}
The coefficient of the divergence is precisely the same as the late-time growth rate of the action in the dynamical cases. 

This is a nice result, which makes a close connection between the complexity calculation for typical states and dynamical calculations. However, it depended on a choice for the WdW patch which might seem more in keeping with the letter than the spirit of the original proposal. The alternative choice for the WdW patch could also offer a sharper probe of the proposed geometry, as it extends into the left region. 

Let us therefore consider the calculation of the action for the patch on the right in figure \ref{wdw}. There is some cutoff in the left region at $r=r_L$. We will also cut off the integration over $t$ by introducing null sheets extending from $r=r_L, t = \pm t_{max}$ into the black hole. We are again interested in the contributions which diverge as $t_{max} \to \infty$. There will be divergent contributions from the bulk integral in the black hole and white hole regions, and from the GHY surface term along the future and past singularities. The divergences with $t_{max}$ for each of these terms are the same as in the previous calculation, so we now get twice the previous result, from taking into account the past and future regions, 
\begin{equation} 
I_{bulk}^{F,P} + I_{surf}^{F,P}  = \frac{\Omega_{d-1}}{8\pi G} (d \mu - 2 \frac{r_+^{d}}{\ell^2})  t_{max} + \ldots
\end{equation}
There is also a divergence from the bulk integral in the left region, 
\begin{equation} 
I_{bulk}^{L}   = -  \frac{\Omega_{d-1}}{4\pi G \ell^2} ( r_L^d - r_+^d)  t_{max} + \ldots 
\end{equation}

Any surface term at the left boundary will also produce a divergent contribution proportional to $t_{max}$, so the calculation now depends on our model for this boundary. One possibility would be to imagine that the spacetime closes off smoothly here, as in the AdS soliton; there would then be no boundary contribution, as in \cite{Reynolds:2017jfs}, and the action would be 
\begin{equation} 
I   =  \frac{\Omega_{d-1}}{8\pi G} ( -2 \frac{r_L^d}{\ell^2}  + d \mu)  t_{max} + \ldots 
\end{equation}

However, this is a problematic result, as the value could be negative if $r_L$ is big enough, or if $\mu$ is small for $k=-1$ (as $r_L > r_+$, and for $k=-1$ $\mu \to 0$ at finite $r_+$). Therefore, we should probably think about the left boundary as more like an end of the world brane, as in \cite{Kourkoulou:2017zaj}. Then we should at least include a GHY term at $r=r_L$. The GHY surface term is
\begin{equation} 
I_{surf}^L = - \frac{r_L^{d-1}}{8 \pi G}  \Omega_{d-1} \left(  \partial_r f (r_L)+ \frac{d-1}{r_L} f(r_L) \right)  t_{max}. 
\end{equation}
Adding this contribution, 
\begin{equation} 
I  =  \frac{\Omega_{d-1}}{4\pi G} (d-1) ( \frac{r_L^d}{\ell^2}  + k r_L^{d-2} )  t_{max} + \ldots 
\end{equation}

This result has, oddly, no direct dependence on the mass of the black hole. The smallest possible value is when $r_L = r_+$. Then $I= 4 M t_{max} + \ldots$; we get precisely twice the divergence in the first WdW patch calculation, as expected since the calculation includes two regions behind the horizon, one in the past and one in the future. The coefficient of the divergence increases monotonically as we increase $r_L$, moving the breakdown of the geometric description into the left region. 

The larger coefficient of the divergence means this calculation has a less direct connection with the previous dynamical studies, but it is not a serious problem; there is no dynamical process here, so there's no reason to expect a bound on the coefficient similar to the Lloyd bound \cite{lloyd}. Cutting off this divergence could still lead to the expected maximal value of the complexity if the geometry breaks down at some time scale $t \sim e^{S}$, the coefficient in this relation just needs to be smaller. Thus, this calculation is still reasonably consistent with expectations for typical states; if this were the correct prescription for CA calculations, the proposed geometry would still be a credible dual for typical states. However, we might expect the coefficient of the divergence to at least be related to the black hole parameters, rather than being set by some arbitrary UV scale. So from the point of view of this calculation, it might seem natural to take $r_L$ of the order of $r_+$, rather than a fixed large value as in \cite{deBoer:2018ibj,deBoer:2019kyr}. 

In summary, including the full region behind the horizon in the geometric description of typical states produces a classical divergence in the complexity that does seem to match generic expectations for typical states. In the fuzzball or firewall proposals, one could imagine that the exponentially large value of the complexity for typical states could be reproduced instead by a classically divergent contribution from the singular structure which replaces the horizon. This is certainly possible; particularly for the action one can imagine getting such a divergent contribution (although divergences in the action for the bulk description of individual microstates could be problematic for other arguments, for example about tunneling rates \cite{Mathur:2008kg}). However, it is encouraging that the geometric description for typical states proposed in \cite{deBoer:2018ibj,deBoer:2019kyr}  naturally produces such a divergence, in a way which is consistent with our heuristic ideas about the relation of complexity to the growth of the wormhole geometry in the spacetime. 

\section*{Acknowledgements}
This work was supported in part by STFC under consolidated grant ST/P000371/1. 

\bibliographystyle{JHEP}
\bibliography{complexity}

\end{document}